\documentstyle[12pt,aasms4]{article}
\lefthead{Dokuchaev, Eroshenko \& Ozernoy}
\righthead{Gamma Ray Bursts from the Evolved Galactic Nuclei}
\begin{document}
\begin{center}
{\sl To appear in the Astrophysical Journal}
\end{center}
\vspace{0.5 truein}
\title {Gamma Ray Bursts from the Evolved Galactic Nuclei}
\author{V. I. Dokuchaev, Yu. N. Eroshenko}
\affil{Institute for Nuclear Research of Russian Academy of Sciences, \\
60th October Anniversary Prospect 7a, Moscow, 117312 Russia;
dokuchaev@inr.npd.ac.ru}
\and
\author{L. M. Ozernoy}
\affil{Physics \& Astronomy Department \\and Institute for Computational
Sciences \& Informatics, George Mason University\\
also Laboratory for Astronomy and Solar Physics, NASA/GSFC;\\
ozernoy@science.gmu.edu, ozernoy@stars.gsfc.nasa.gov}
\begin{abstract}
A new cosmological scenario for the origin of gamma ray bursts (GRBs) 
is proposed. In our scenario, a highly
evolved central core in the dense galactic nucleus is formed 
containing a subsystem of compact stellar remnants (CSRs), such as
neutron stars and black holes.  Those subsystems result from the
dynamical evolution of dense central stellar clusters in the galactic
nuclei through merging of stars, thereby forming (as has been realized by 
many authors) the short-living massive stars and then CSRs. We
estimate the rate of random CSR collisions in the evolved galactic
nuclei by taking into account, similar to Quinlan \& Shapiro (1987),
the dissipative encounters of CSRs, mainly due to radiative losses of
gravitational waves, which results in the formation of intermediate
short-living binaries, with further coalescence of the companions to
produce GRBs.
We also consider how the possible presence of a central supermassive black
hole, formed in a highly evolved galactic nucleus, influences the
CSR binary formation.
This scenario does not postulate {\it ad hoc}\/ a required number of
tight binary neutron stars in the galaxies. Instead, it gives, for
the most realistic parameters of the evolved nuclei,
the expected rate of GRBs consistent with the observed one, 
thereby explaining the GRB appearance in a
natural way of the dynamical evolution of galactic nuclei. In
addition, this scenario provides an opportunity for a cosmological
GRB recurrence, previously considered  to be a distinctive
feature of GRBs of a local origin only. We also discuss some other
observational tests of the proposed scenario.
\end {abstract}
\keywords{gamma-rays: bursts --- galactic nuclei --- black hole physics}
\newpage
\section{Introduction}
In spite of significant progress in the accumulation of data on
GRBs they are still mysterious --- chiefly because of uncertainty
whether they represent a local or a cosmological phenomenon
or a mixture of both
(for reviews, see e.g. Meegan et al. 1992, Lamb 1995, Paczy\'nski 1995)
{\it and} due to the enigmatic nature of the central engine.

If GRBs are cosmological in origin (Usov and Chibisov 1975; Mao and
Paczy\'nski 1992), they have to have a huge intrinsic luminosity --- on
the level of, or higher than, the luminosity of a supernova explosion.
In the framework of a cosmological scenario, different possible sources
of GRBs have been proposed, such as the collapse of supermassive
magnetic stars (Usov and Chibisov 1975; Ozernoy and Usov 1973), the merging
of neutron stars (NSs) or neutron stars and black holes (BHs) in close
binaries of distant galaxies (Cavallo and Rees 1978; Paczy\'nski 1986;
Goodman 1986), the superconducting cosmic strings (Plaga 1994), etc.

If at least a part of GRBs occurs in the
Milky Way halo, they are generated in
the magnetosphere or on the surface of NSs (Higdon and Lingenfelter
1990; Harding 1991). Isotropy of GRB distribution puts severe
constraints on both the distribution of the relevant NSs between the
Galactic disk and halo and the associated  physical parameters of NS
(Paczy\'nski 1991; Li \& Dermer 1992; Gurevich et al. 1993; Hakkila et
al. 1995; Hartmann \& Narayan 1996; Briggs et al. 1996).

Owing to statistical uncertainties and inhomogeneities of current GRB
samples, available data on possible scalings of bright and dim GRB
durations, fluxes, and spectra fail to provide an unambiguous
distinction between the cosmological and local (the Ga\-lac\-tic) origin of
GRBs.  Some statistical investigations of different GRB distributions
give support to cosmological location of GRBs (Norris et al. 1994;
Wijers \& Paczy\'nski 1994; Kolatt \& Piran 1996), whereas others
favor the local one (Gurevich et al.  1993; Hakkila et al. 1995;
Chernenko \& Mitrofanov 1995; Mitrofanov et al.  1995).

In the framework of a cosmological scenario, the fitting of the integral 
GRB luminosity function [i.e. source number $N(F)$
with photon fluxes $>F$] for different GRB samples and model
assumptions about the source spectra and evolution makes it possible to
estimate the GRB peak luminosity and the production rate of GRBs per
comoving unit volume to be, correspondingly  (Mao  \& Paczy\'nski
1992; Cohen \& Piran 1995; M\'esz\'aros \& M\'esz\'aros 1995; Rutledge
et al. 1995)
\begin{equation}
L_{\gamma}\sim10^{51} - 10^{52}\mbox{~erg}\mbox{~s}^{-1},\label{Lpeak}
\end{equation}
\begin{equation} ~
\dot n_{0}\sim10^{-8}-10^{-7}\mbox{~events}\mbox{~yr}^{-1}
\mbox{~Mpc}^{-3}. \label{nrate}
\end{equation}
Eq.~(\ref{nrate}) implies that the rate of NS merging in galaxies
necessary to explain the GRB phenomenon must be of the order of
\begin{equation}
\dot N_{g}\sim10^{-6}-10^{-5}~\mbox{events
~yr}^{-1} ~\mbox{galaxy}^{-1}.  \label{Ngrate}
\end{equation} 
This value of $\dot N_{g}$ is consistent with the calculations of the 
NS merging rate for the Galactic binaries (Clark et al. 1979; 
Narayan et al. 1991):  
$\dot N_{g}\sim10^{-6}-10^{-4}~\mbox{events~yr}^{-1}$ per galaxy (see
however Sec.~4).

It can be found from Eq.~(\ref{Lpeak}) that in each act of NS merging
the energy $Q_{\gamma}\sim10^{51}-10^{52}$~erg is emitted (assuming
isotropic radiation) in the form of gamma rays. Therefore the
transformation efficiency of the total energy $Q$ of two merging NSs
into gamma rays must be of the order of $\eta=Q_{\gamma}/Q\sim10^{-3}-10^
{-2}$. This value being rather
large is challenging for the models of a GRB generation by NS merging.
In the concept of relativistic fireball as radially expanding, optically
thick electron-positron plasma (Cavallo and Rees 1978; Paczy\'nski 1986; 
Goodman 1986; Shemi and Piran 1990), the value of $\eta$ can be made 
smaller by taking into account the
possible relativistic boosting of the observed GRB intensity if the
fireball expansion occurs with the bulk Lorentz-factor
$\Gamma\sim10^{2}-10^{3}$ (Yi 1993; Yi \& Mao 1994). However, the
necessary rate of GRB generation in the unit comoving volume would also
be increased by a factor of $ (2\Gamma)^{2}$ thereby destroying the
above consistency between the inferred and estimated $\dot N_g$.

This problem is abandoned in a different cosmological scenario
proposed here: a GRB generation is considered to occur
by involving coalescence of binary NSs and/or
BHs {\it  formed by radiative capture in the central compact stellar
clusters of the evolved galactic nuclei}. This new scenario of GRB
origin is complementary to the standard cosmological scenario, in
which coalescence of neutron stars in very distant galaxies is
assumed, without specifying the origin of parent binaries. In Sec.~2,
dynamical evolution of galactic nuclei leading to the formation of
the central dense stellar clusters is briefly outlined. Sec.~3 deals with
the rate of collisions and coalescence of compact stars which result in GRB
production. Discussion of our scenario is given in Sec.~4.
A brief account of this work was published elsewhere
(Ozernoy, Dokuchaev \& Eroshenko 1996).

\section{Central Compact Clusters in the Galactic Nuclei}
According to well known models of the dynamical evolution of stellar
systems, those galactic nuclei which are dense enough, inevitably
proceed through the stellar collision and coalescence  stage (Spitzer
and Saslaw, 1966; Colgate 1967; Sanders 1970; Spitzer 1971; Dokuchaev
1991).  This process is enhanced by the hardening of hard binaries formed
in several ways (Ozernoy \& Dokuchaev 1982, Dokuchaev \& Ozernoy
1982). The enhanced evolution of the galactic nucleus results in a
copious formation of short-living massive stars and subsequent
supernova explosions implying the `build up' of a subcluster of
compact stellar remnants (CSRs) with an appreciable  fraction of NSs
and BHs (Begelman \& Rees 1978, Quinlan \& Shapiro 1990 and
refs. therein).  Since so formed binaries are much heavier than the
field stars, they settle on a relaxation time scale (which is much
shorter than the cluster's age) to the center of the nucleus, thus
making the subcluster of CSRs contain an enhanced fraction of
tight binaries (Dokuchaev \& Ozernoy 1982, Quinlan \& Shapiro 1990).

Let us consider a central stellar cluster in the galactic nucleus of
mass $M$ and radius $R$, consisting of $N\gg 1$ CSRs, the relics of massive
stars.
The CSR velocity dispersion in the cluster is
given by the virial theorem to be $v\simeq(GM/2R)^{1/2}$. If the
space density of galaxies with the evolved galactic nuclei contributes
a fraction $\Omega_{g}$ of the critical density
$\rho_{c}=3{H_{0}}^{2}/8\pi G$, where
$H_{0}=75\,h$~km~s$^{-1}$~Mpc$^{-1}$ is the Hubble constant, then the
mean number density of the evolved galactic nuclei in the Universe is
\begin{equation}
n_{g}=\rho_{c}\Omega_{g}/M_{g}=1.6\cdot 10^{-2}\,h^{2}
\left(\frac{\Omega_{g}}{10^{-2}}\right)
\left({M_g\over 10^{11}\,{\rm M}_{\sun}}\right)^{-1}~{\rm Mpc}^{-3},
\label{ng}
\end{equation}
assuming the evolved galactic nuclei to be the typical inhabitants
of the bulk of galaxies,
$M_{g}$ being the luminous mass of the typical host galaxy.
Below, we find the allowed range of parameters $M$ and $R$, for
which encounters/collisions of CSRs in the host galactic nuclei can provide
the observed GRB rate.
We also take into account the possible presence of a central supermassive
black  hole formed in a highly evolved galactic nucleus and we consider
how its presence influences the CSR binary formation.
\section{Collisions and Coalescence of Compact Stellar Remnants}
\subsection{Radiative Capture of CSRs vs. Direct Collision}
The cross section for coalescence of two identical CSRs of mass $m$
moving with a relative velocity $v_{\infty}=\sqrt2\,v$ can be presented,
similarly to the cross section for the capture of nonrelativistic test
particles by a massive object, in the form:
\begin{equation}
\sigma_{coll}=\pi r_{p}^{2}{\left(1+\frac{r_{g}}{r_{p}}
\frac{c^{2}}{v^{2}}\right)}\approx\pi r_{p}r_{g}
{\left(\frac{c}{v}\right)}^{2},\label{sigmacoll}
\end{equation}
where $r_{g}=2Gm/c^{2}$ is the gravitational radius of a CSR, and
$r_{p}$ is the maximum collision periastron separation allowing for 
CSR coalescence. For the direct coalescence of two CSRs,
$r_{p}=2r_{g}$ (Shapiro \& Teukolsky 1985).

We also take into account the formation of compact star binaries
through dissipative two-body encounters via gravitational radiation.
The maximum periastron separation for the radiative capture of
compact stars which results in the formation of a hard binary (i.e. a
binary of a binding energy $\epsilon_{b}\gtrsim mv^{2}/3$) is
estimated to be $r_{p}\approx(3/2)r_{g}(c/v)^{4/7}$ (see Giersz 1985;
Quinlan \& Shapiro 1987 for details). Therefore, the cross
section for the radiative capture into a binary is given by
\begin{equation}
\sigma_{cap}\approx\frac{3}{2}\pi
r_{g}^{2}{\left(\frac{c}{v}\right)}^{18/7}. \label{sigmacap}
\end{equation}
We define the BH collision as such a close encounter of two
BHs, which leads to the intersection of their event horizons so as to
inevitably terminate in their coalescence.
The total energy radiated during the head-on collision of two
identical Schwarzschild BHs on a parabolic orbit is
$\Delta\epsilon_{gr}\simeq2.5\cdot10^{-3}mc^{2}$ (Smarr 1979).
Therefore, the radiative capture of two CSRs into a binary can take place
in the cluster with velocity dispersion
\begin{equation}
v\lesssim v_{cap}=(\Delta\epsilon_{gr}/m)^{1/2}\simeq1.5\cdot10^{4}
\mbox{~km~s}^{-1}.
\label{velocap}
\end{equation}
In clusters with $v\gtrsim v_{cap}$, gravitational radiation losses during
two-body CSR encounters are insufficient for the formation of binaries.
Meanwhile in a nonrelativistic cluster with $v\lesssim v_{cap}$,
as can be seen from Eqs.~(\ref{sigmacoll})--(\ref{velocap}),
the capture cross section
$\sigma_{cap}$ of radiative binary formation exceeds
the direct collision cross section  $\sigma_{coll}$  by a large factor
$r_p^{cap}/r_p^{coal}\simeq(3/4)(c/v)^{4/7}$.

Components of a circular binary of radius $r$  coalesce, due to
successive gravitational radiation losses, in a time
\begin{equation}
t_{gr}=\frac{5}{64}\frac{r_{g}}{c}
{\left(\frac{r}{r_{g}}\right)}^{4}.\label{timegrav}
\end{equation}
The bulk of the newly formed tight binaries are highly
eccentric, $1-e\ll 1$. In this case, the exact solution for time evolution
of eccentric binary orbital parameters
 due to the quadrupole gravitational radiation (Pierro \& Pinto 1996)
indicates that the binary lifetime is of
the same order as $t_{gr}$ given by Eq.~(\ref{timegrav}) with $r=r_p$,
where $r_{p}\approx(3/2)r_{g}(c/v)^{4/7}$ is the periastron
separation during the first encounter. This has a simple physical
meaning: circularization of a highly eccentric orbit is fast compared to
the subsequent slow
evolution of a nearly circular orbit until two CSR components coalesce.
The value of $t_{gr}$ is less than the cosmological time [say, the age
of the flat Universe $t_{H}=(2/3)H_{0}^{-1}$] in the galactic nuclei
with velocity dispersion
\begin{equation}
v\gtrsim v_{H}=c{\left(\frac{5\cdot3^{5}}{2^{11}}
\frac{H_{0}}{c}r_{g}\right)}^{7/16}.\label{velocity}
\end{equation}
The numerical value of the r.h.s. is astonishingly low:
$v_{H}\simeq3.0(m/{\rm M_{\sun}})^{7/16}$~cm~c$^{-1}$, far less of any real
velocity dispersion in the central stellar cluster. This implies that
actually
there is no lower limit to velocity dispersion of CSRs able to form
binaries by radiative capture: all
such formed binaries are short-living compared to the
age of the Universe, i.e. they have more than enough time to coalesce.
Below, only those central clusters in the galactic nuclei are considered
for which the upper limit to $v$ given by Eq.~(\ref{velocap}) is satisfied.
\subsection{The Rate of CSR Radiative Capture}

In this subsection, we mainly follow  to Quinlan \& Shapiro (1987,
1989) to calculate the  CSR collision rate in an evolved stellar
cluster. The rate of CSR radiative capture into a binary, followed by 
a successive coalescence of the components and then a GRB event, is given 
by $\dot N_{c}\simeq(1/2)Nn\sigma_{cap}v_{\infty}$ per one galactic nucleus,
where $n$ is the CSR mean number density and
$N=M/m\simeq(4\pi/3)R^{3}n$ is the total number of CSRs in the
central cluster. As a result, the rate of CSR collisions is given by
\begin{equation}
\dot N_{c}\simeq9\sqrt2{\left(\frac{v}{c}\right)}^{17/7}
\frac{c}{R} \simeq 5.8\cdot10^{-6}\,
{\left(\frac{M}{10^7\,{\rm M}_{\sun}}\right)}^{17/14}
{\left(\frac{R}{0.1\,{\rm pc}}\right)}^{-31/14}
\mbox{~events~yr}^{-1}\mbox{~galaxy}^{-1},\label{Nrate}
\end{equation}
where fiducial parameters for the central nuclear cluster
$M=10^7\,{\rm M}_{\sun}$ and $R=0.1$~pc are used (see discussion in
Sec.~4). Therefore, the rate of GRBs per unit of comoving volume is expected
to be
\begin{eqnarray}
\dot n_{0}&=&\dot N_{c}n_{g}\simeq\frac{27}{8\pi\sqrt2}\,\Omega_{g}
{\left(\frac{H_{0}}{c}\right)}^{2}{\left(\frac{v}{c}\right)}^{3/7}
\frac{M}{M_{g}}\frac{c}{R^{2}} \nonumber\\
&\simeq & 9.1\cdot10^{-8}\,
{\left(\frac{\Omega_{g}{h^{2}}}{10^{-2}}\right)}
{\left(\frac{M_g}{10^{11}\,{\rm M}_{\sun}}\right)}^{-1}
{\left(\frac{M}{10^7~{\rm M}_{\sun}}\right)}^{17/14}
{\left(\frac{R}{0.1\,{\rm pc}}\right)}^{-31/14}\:
\frac{\mbox{events}}{\mbox{yr~Mpc}^{3}}.\label{Nrategal}
\end{eqnarray}
Eqs.~(\ref{Nrate}) and (\ref{Nrategal}) are fairly consistent with
the rate of NS merging in galaxies required to explain the GRB
phenomenon as mentioned in the Introduction [see Eqs. (\ref{Ngrate}) and
(\ref{nrate})].  Yet, while confronting the
value of $\dot n_0$ given by Eq.~(\ref{Nrategal}) with Eq.~(\ref{nrate}),
one should
emphasize that such a comparison would not solely depend on CSR
merging. In fact, it would incorporate two more
factors:  (i) evolutionary effects associated with how the CSR merging
rate evolves in time as a result of evolution of galactic nuclei, and
(ii) effects depending on the choice of a cosmological model
(parameters $q_0$, $H_0$, etc.).  Each specific model for galactic
nucleus evolution would come up with a corresponding relationship
$M=M(R)$. This issue is out of the scope of the present paper.
\subsection{An Allowed Range of Masses and Radii
of the Evolved Galactic Nuclei as GRB Sources}
Here, we constrain ourselves by evaluating the range of parameters of a
central dense cluster in which coalescence of NSs or BHs could
provide the rate of GRBs $\dot n_{0}$ consistent with the observed
one.  By incorporating the numerical estimation for the inferred
value of $\dot n_{0}$ given by Eq.~(\ref{nrate}) [or,
correspondingly, $\dot N_g$ given by Eq.~(\ref{Ngrate})], we find from
Eq.~(\ref{Nrategal}) the relationship of interest between $M$ and
$R$:
\begin{equation}
M=M_{GRB}\simeq1.6\cdot10^{6}\,
{\left(\frac{\Omega_{g}}{10^{-2}}
\frac{h^2}{\dot n_{-8}M_{g\,11}}\right)}^{-14/17}
{\left(\frac{R}{0.1\,{\rm pc}}\right)}^{31/17}\,{\rm M}_{\sun},
\label{M}
\end{equation}
where $\dot n_{-8}=\dot n_{0}/10^{-8} \mbox{~events}\mbox{~yr}^{-1}
\mbox{~Mpc}^{-3}$ and $M_{g\,11}=M_{g}/10^{11}~{\rm M_{\sun}}$.

It is instructive to discuss available constraints on the range of the
allowed $M$ and $R$. One of the general physical constraints
is the requirement that the cluster radius be
greater than several gravitational radii, say, $R\gtrsim 5R_{g}=10\,GM/c^{2}$
(Zel'dovich \& Podurets 1965).  Obviously, this is a very stringent
condition and it would be more realistic for the time of
dynamical evaporation of the central cluster, $t_{ev}\simeq40\,t_{r}$,
where $t_{r}$ is the two-body relaxation time (Spitzer \& Hart 1971),
to marginally exceed the age of the universe:
\begin{equation}
t_{ev}\simeq\frac{10}{\pi}\frac{v^{3}}{G^{2}m^{2}n\ln(0.4N)}
\gtrsim\frac{2}{3}\,H_{0}^{-1}\label{tevap}
\end{equation}
(otherwise, i.e. in the case of $t_{ev}\ll H_0^{-1}$, there would be an
inappropriately high recurrence of GRBs due
to an overabundant CSR merging in a too dense central core). Another
constraint to this model is a firm (though a trivial one) upper limit
to the mass of the nucleus compared to that of the host galaxy:
 $M\lesssim M_{g}$. Although a more sophisticated relationship between
$M$ and $M_{g}$ depends on (yet unknown) conditions of galaxy formation,
a more realistic upper limit to $M$ consistent with the available data
would be $M\lesssim 10^{-3}~M_{g}$ or something of that nature.

Eqs.~(\ref{M}), (\ref{tevap}), and the condition $M\lesssim M_{g}$
yield an allowed range of values for radius $R$ and mass $M$ of the
central compact cluster [we put $\ln(0.4N)=10$ throughout the
numerical estimations]:
\begin{equation}
0.41\,h^{-3/41}
{\left(\frac{\Omega_{g}}{10^{-2}\dot n_{-8}}\right)}^{7/41}
{\left(\frac{m}{\rm M_{\sun}}\right)}^{17/41}M_{g\,11}^{-7/41}
\lesssim\frac{R}{1\,{\rm pc}}\lesssim45\,
{\left(\frac{\Omega_{g}h^{2}}{10^{-2}\dot n_{-8}}\right)}^{14/31}
M_{g\,11}^{3/31};\label{Rrange}
\end{equation}
\begin{equation}
2.1\cdot10^{7}\,h^{-73/41}
{\left(\frac{\Omega_{g}}{10^{-2}\dot n_{-8}}\right)}^{-21/41}
{\left(\frac{m}{\rm M_{\sun}}\right)}^{31/41}
M_{g\,11}^{21/41}\lesssim\frac{M}{{\rm M}_{\sun}}
\lesssim10^{11}\,M_{g\,11}.\label{Mrange}
\end{equation}
Hence, Eqs.~(\ref{Rrange}) and (\ref{Mrange}) define, in the framework
of our cosmological scenario, a range of radius $R$ and mass $M$ [connected
by relationship ~(\ref{M})] for those dense CSR clusters, which
could provide, by the CSR coalescence, the necessary GRB rate and fluence
consistent with observations. The most appropriate values of $R$ and
$M$ for the evolved galactic nuclei scenario would correspond to the
left-hand-sides of Eqs.~(\ref{Rrange}) and (\ref{Mrange}) where
$t_{ev}=t_{H}$, i.e.  $R\approx0.4$~pc and $M\approx10^{7}~{\rm M_{\sun}}$,
respectively.

The total emitted energy of two colliding stellar mass BHs (which
considerably exceeds that of two NS collision) is radiated primarily in
the form of strong gravitational waves. Since each colliding BH
supposedly contains an accretion disk/torus (AD), left since the collapse of
the massive predecessor, coalescence of two BH+AD systems should
evidently be accompanied by a generation of hard radiation, presumably
gamma rays, although the transformation efficiency of
available energy into  gamma rays is uncertain until making rather
complicated computations.  Yet a high transformation efficiency (on
the level of, or higher than, that of NS-NS merging) seems to be
reasonable if a fireball is formed in the process of two BH+AD
collisions.  Some small amount of matter in an AD around one or both
colliding BHs can act as a ``primer'' to ignite a fireball. The AD
matter can be gradually piled up from the ambient
gas by the moving BH.

Although the parameters of gamma rays from two colliding BH+AD systems
could only be found in further work, this does not really influence our
basic conclusions since
the fraction of massive stellar BHs in the galactic nuclei seems to be
smaller (perhaps even much smaller) than that of NSs. Therefore, the bulk
of GRB progenitors is expected to be related with NS--NS and NS--BH, rather
than BH--BH, collisions.
\subsection{Influence of a Central Massive Black Hole}
In the above consideration, an evolved core of the galactic nucleus
is assumed to be not too far evolved, in the sense that a central massive 
BH has not yet had time to be formed, or its mass is still negligible.
In this subsection, we consider the opposite case, when an already formed
central BH has acquired a large enough mass to
substantially influence the dynamics of the surrounding stellar core.
Specifically, we assume that an evolved galactic nucleus harbors
a supermassive black hole
(SMBH) of mass $M_h\gtrsim 10^{7} - 10^{8}{\rm M}_{\sun}$
embedded into a dense cluster of $N$ compact stellar remnants (CSRs),
mostly neutron stars and stellar-mass black holes.
 For simplicity, we assume the
total mass of this cluster $M=mN\ll M_h$ so that the cluster represents
 a star `atmosphere' of radius
$R\simeq GM_{h}/v^{2}$ around the dominating central SMBH,
$v\ll c$ being the  CSR velocity
dispersion in the gravitational field of the latter.

The star `atmosphere' around a SMBH is genetically related to its
precursor, {\it viz.}, a central compact
stellar cluster, whose evolutionary time scale $t_{ev}$ was much less
than the Universe's age $t_H$.
If the SMBH formed as a result of the dynamical evolution of the cluster,
this happened on an evaporational time scale, i.e.
$t_{ev}\simeq40t_r$, where $t_r$ is the two-body relaxation time.
After the formation of the central SMBH in the stellar cluster,
its interaction with surrounding stars, such as tidal
disruption and consumption of stars by the hole,
results in an increase of the evolutionary time scale, until $t_{ev}$
reaches $t_H$  and becomes frozen thereafter. Therefore,
the most probable stage
in which one could find an evolved stellar cluster around a
`dormant' SMBH would be a star `atmosphere' around it with
$t_{ev}\simeq t_H$, which makes the radius of such `atmosphere' to
be 
\begin{eqnarray} 
R&\simeq& 5.75\cdot10^{-2}h^{-2/3}{\left(\frac{m}{\rm M_{\sun}}\right)}^{4/3}
{\left(\frac{N}{10^{6}}\right)}^{2/3}
{\left(\frac{M_{h}}{10^{7}{\rm M_{\sun}}}\right)}^{-1}{\mbox{~pc}}.
\label{R2}
\end{eqnarray}
This yields $R\gtrsim10^{-2} - 10^{-3}$~pc if $N\sim10^{6} -10^{7}$ and
$M_h\gtrsim 10^{7} - 10^{8}\,{\rm M}_{\sun}$. 

In the dense star `atmosphere' around the central SMBH, close
encounters of two CSRs result in a radiative capture followed by a
successive coalescence and a GRB event. The capture rate is given, as before,
by $\dot N_{c}\simeq(1/2)Nn\sigma_{cap}v_{\infty}$, but its value
\begin{equation}
\dot N_{c}\simeq\frac{9}{2\sqrt2}{\left(\frac{Nm}{M_{h}}\right)}^{2}
\left(\frac{v}{c}\right)^{17/7}\frac{c}{R}\label{NBHrate}
\end{equation}
differs from that given by eq. (10) by a factor
${\left({Nm}/{M_{h}}\right)}^{2}$
since $v$ now depends upon the BH mass.

Assuming that the number density of `dormant' SMBHs is
about the same as that of giant galaxies,
$n_{g}\sim10^{-2}~{\mbox{Mpc}^{-3}}$,
the GRB rate  per unit volume of the Universe is expected to be
\begin{eqnarray}
\dot n_{0}&\!\!\!\!=&\!\!\!\!\dot N_{c}n_{g}
\nonumber\\
&\!\!\!\!\simeq&\!\!\!\!2.7\cdot10^{-8}
{\left(\frac{n_g}{10^{-2}\,\mbox{Mpc}^{-3}}\right)}
{\left(\frac{M_{h}}{10^7\,{\rm M}_{\sun}}\right)}^{-\frac{11}{14}}
{\left(\frac{Nm}{10^6{\rm M}_{\sun}}\right)}^{2}\!
{\left(\frac{R}{0.01\,{\rm pc}}\right)}^{-\frac{31}{14}}
\frac{\mbox{events}}{\mbox{yr~Mpc}^{3}}.\label{nrategalBH}
\end{eqnarray}
One can see that this rate is consistent with the inferred cosmological
GRB rate given by equation (2)
if the radius of the SMBH star ``atmosphere'' is as small as
\begin{equation}
R\simeq2.2\cdot10^{-2}
{\left(\frac{\dot
n_{GRB}}{10^{-8}\,\mbox{yr}^{-1}\mbox{Mpc}^{-3}}\right)}^{-\frac{14}{31}}
{\left(\frac{n_g}{10^{-2}\,\mbox{Mpc}^{-3}}\right)}^{\frac{14}{31}}
{\left(\frac{M_{h}}{10^7\,{\rm M}_{\sun}}\right)}^{-\frac{11}{31}}
{\left(\frac{Nm}{10^6{\rm M}_{\sun}}\right)}^{\frac{28}{31}}
{\mbox{~pc}}.\label{R1}
\end{equation}
This radius is consistent with what is expected according to equation
(\ref{R2}) for an evolved stellar core with a central SMBH.

\subsection{Recurrence of Cosmological GRBs}
Some authors consider a would-be-found GRB recurrence as serious evidence
for the origin of GRBs in the Galactic halo (see e.g. Bennett \& Rhie 1996;
Tegmark et al. 1996). However, as we demonstrate below, GRB recurrence
could be expected in the cosmological case as well. Eq.~(\ref{M})
represents a relationship only between the {\it average} radius $R$
and mass $M_{GRB}(R)$ of the evolved galactic nuclei able to provide the 
observed rate of GRBs.  The actual parameters of the evolved nuclei can 
be widely spread around the average ones, and some nuclei may be very
compact. Our model indicates that an extremely compact nucleus  would be a
source of multiple (recurrent) cosmological GRBs. If the influence
of the forming central SMBH is negligible, the radius of the
cluster $R$ in which the CSR coalescence rate is $\dot N_c$ is given,
according to Eq.~(\ref{Nrate}), by
\begin{equation} R\simeq
3.6\cdot10^{-5}\, {\left(\frac{M}{10^7\,{\rm M}_{\sun}}\right)}^{17/31}
{\left(\frac{\dot N_{c}}{10^2\,{\rm yr}^{-1}}\right)}^{-14/31}
\mbox{~pc}. \label{Rrec}
\end{equation}
The corresponding velocity dispersion $v$ in the cluster is
\begin{equation}
v\simeq 2.4\cdot10^{4}\,
{\left(\frac{M}{10^7\,{\rm M}_{\sun}}\right)}^{7/31}
{\left(\frac{\dot N_{c}}{10^2\,{\rm yr}^{-1}}\right)}^{7/31}
\mbox{~km}\mbox{~s}^{-1}, \label{vrec}
\end{equation}
The characteristic evolutionary time of a cluster as dense as this is
determined by the CSR capture time (Quinlan \& Shapiro, 1987):
\begin{equation}
t_{cap}=\frac{N}{\dot N_{c}} \simeq 5.8\cdot10^{3}\,
{\left(\frac{M}{10^7\,{\rm M}_{\sun}}\right)}^{-3/14}
{\left(\frac{R}{10^{-5}\,{\rm pc}}\right)}^{7/31} \mbox{~yrs},
\label{tcap}
\end{equation}
i.e. this evolutionary stage is very brief.
Recurrent GRBs might be also associated with a much more advanced stage
when the gravitational field is dominated by a central SMBH.
The rate of CSR coalescence, 
accounting for the SMBH influence, is evaluated elsewhere.

After this paper was basically completed
and its brief account, with mentioning of the possibility of recurrent GRBs
was published (Ozernoy, Dokuchaev \& Eroshenko 1996), we become aware of
the BATSE observations of 4 GRB events during two days
in October 1996 from the same direction in the sky (Meagan et al. 1997).
Unless this is just one unusually long event, this might be  explainable 
in the framework of our model as  the CSR coalescence in a
highly evolved cluster of radius $R\approx10^{-5}$~pc
and mass $M\approx10^7{\rm M_{\sun}}$, as Eq.~(\ref{Rrec}) indicates.
This stage of cluster evolution,
which only lasts $t_{cap}\approx10^3-10^4$~yrs,
might precede the subsequent formation
of a central massive BH, although its mass at this stage is much smaller
than the cluster mass $M$.
An alternative possibility to explain such a fast GRB recurrence
would be CSR coalescence in the vicinity of a forming (or an already formed)
central massive BH. We explore both these possibilities elsewhere
in more detail, including possible reasons for the apparent
absence of GRBs, both before and after the recurrent events were detected.

\section{Discussion}

A recent Beppo-Sax finding of distant galaxies as counterparts for two GRBs 
(e.g. Sahu et al. 1997,  Metzger et al. 1997) has confirmed that at least 
a part of the gamma ray bursts (GRBs) have a cosmological location. Thus,  
our new scenario for the origin of GRB progenitors proposed in this paper 
stimulates and challenges its further development and testing.

A novel aspect of the evolved galactic nuclei explored above
is that those nuclei, which contain in
their dense central parts numerous compact stellar remnants
(CSRs) such as neutron stars (NSs) and black holes (BHs), might be
appropriate sites for the production of GRBs. Collisions
between those remnants can achieve, in the most natural
way, a high rate of GRBs. A similar model has been proposed for
GRB origin in the hypothetical halo dark clusters (Carr \& Lacy 1987,
Wasserman \& Salpeter 1994). While there is no conclusive evidence in
favor of those dark clusters, the production of a large number of
tight NS and BH binaries in the galactic nuclei is a natural consequence 
of dynamical evolution of the latter.  Compact galactic nuclei as a site 
for GRBs do not look as speculative as halo  dark clusters. Besides, 
this would enable one to connect the GRB aspect with a previous, and much
more solid, work on dynamics of dense galactic nuclei.
Moreover, there is a mounting evidence in favor of many dark stellar
remnants in the nucleus of the Milky Way Galaxy (Haller et al. 1996,
Lipunov, Ozernoy, Popov et al. 1996), although the present number
density of those remnants is apparently not high enough to produce
GRB events at the Galactic center.

An encouraging point is that the estimations of the allowed ranges for
radii and masses of dense stellar clusters where collisions of CSRs
could result in their coalescence and GRB production
[Eqs.~(\ref{Rrange}) and (\ref{Mrange})], give quite appropriate values
consistent with the observed ranges for the galactic nuclei.  Yet, it
remains to be seen whether the outlined cosmological scenario of
GRB origin actually provides both the observed GRB rate and flux.
Furthermore, the distribution of galactic nuclei in mass $M$ and radius
$R$ has not been taken into account. In reality, the parameters of
galactic nuclei change in the process of the dynamical evolution.
Accordingly, the rate of compact star remnants collisions changes too
and,  by the present time, in many galactic nuclei some part of the
nucleus could collapse or evaporate. Therefore, in the present approach,
the nucleus parameters $M$ and $R$ adopted above must be considered only as
fiducial values.  Generalization of the present analysis with
accounting for possible  distribution of nuclei in mass and radius
would allow to model the observed $\log N(F)$---$\log F$ distribution.

One could argue that the advantage of the standard cosmological
scenario is that the observed statistics of binary pulsars seems to
be consistent with the observed GRB rate (Narayan et al. 1991). Meanwhile
there are arguments that the consistency only exists in a hypothetical,
multi-parametric scenario for compact binary stars evolution (Tutukov
\& Yungelson 1993, Lipunov et al. 1995). In contrast, our
evolved galactic nuclei scenario produces,
as is shown in Sec.~3, an appropriate number of short living binaries fairly
consistent with observations of GRBs and arising naturally as a result of
the dynamical evolution of galactic nuclei.

A relevant point is whether each appropriate galactic nucleus as a GRB 
source has evolved 
so as to produce a central massive black hole (MBH) and to become an active 
galactic nucleus (AGN). As we argue in Sec.~3, GRBs could be produced in the
evolved nuclei both with and without a MBH. Interestingly, a
stage of copious CSR production might even precede the formation of a
supermassive BH. Whether or not the central SMBH actually forms, depends
on a variety of factors (e.g. the formation and hardening of hard binaries),
which are able to either prevent or
retard, stop, and reverse the core collapse. Let us suppose that the
latter happens and results in reversing the core collapse to core
expansion.  Since the post-collapse core evolution slows down around
the time of maximum expansion, observations will find the core, most
probably, to be near that maximum. Fokker-Planck calculations indicate
$r_c/r_h\sim 10^{-2}$ ($r_h$ being the half-mass radius) as a typical
value around the time of maximum expansion (Murphy et al. 1990).
This would justify our choice of $R\sim 0.1-1$~pc while evaluating
$\dot N_c$ with the use of Eq.~(\ref{Nrate}). It is worth noting that
the resulting $\dot N_c$ is fairly consistent with the rate of NS
merging in galaxies required to explain the GRB phenomenon
[Eq.~(\ref{Ngrate})].
The above argument implies that the GRB production might be associated
with a specific stage in the dynamical evolution of galactic nuclei.  The
condition $r_c/r_h\sim 10^{-2}$ can be used to constrain the expected
relationship $M=M(R)$ for the nuclei responsible for GRB production.

We believe that GRB production could be associated with a far more advanced
stage of evolution of galactic nuclei as well, when a central massive black
hole is either forming or has already been formed. In this case, a compact
CSR cluster continues to exist in the vicinity of the MBH and to serve as
a source of GRBs.

Our scenario is, in a sense, inevitable for the evolved galactic
nuclei. At the same time, it is complementary (rather than opposing) to the
standard cosmological scenario. Moreover, in our scenario very hard
(`superelastic') binaries are ejected from the galactic nuclei with
characteristic velocities $\gtrsim 1000$ km/s as a result of interaction
with the field stars in the central stellar cluster [for superelastic
binaries, the ejection velocity is given by
$v_{ej}\gtrsim\left({2\over 3}\cdot 78.75\right)^{1/2}v\approx 724
\left(v/100\,{\rm km~s^{-1}}\right)$ km/s
(Ozernoy \& Dokuchaev 1982, p.3), where $v$ is velocity dispersion].
Therefore it seems possible, at least in principle, to fill the
galactic halo, up to very large distances, with the compact,
short living NS binaries ejected from the galactic nucleus.
However, the core of the Milky Way does not seem be dense enough as to
serve as a source of NS binaries to explain {\it all} the observed GRBs.

In the standard cosmological scenario, a NS+NS binary being a product of
the evolution of the binary's normal stellar constituents, is
expected to be formed with a high (a few $10^2$ km/s) 
kick velocity (e.g. Fryer \& Kalogera 1997). By the time of 
merging and producing a GRB ($\sim 10^8-10^9$ yrs after its birth), 
the binary would be found at a $\sim$ 30 kpc distance from the birthplace.
Therefore, an off-center location of the GRB afterglow found 
at a cosmological distance (Sahu et al. 1997) seems to be consistent
with both new and standard scenarios. It
cannot serve solely as a means of differentiating between the origin of the 
parent binary in the galactic {\it disk} (e.g. in a star-forming region) 
or in the galactic {\it nucleus}.  The host galaxy also could not make
such a differentiation, because the evolved galactic nuclei occur
both in spiral and elliptical galaxies.

In a cosmological GRB source, the formation of an
optically thick fireball of electron-positron plasma (Cavallo and
Rees 1978; Paczy\'nski 1986; Goodman 1986; Shemi and Piran 1990)
is inevitable due to
reaction $\gamma\gamma\to e^{+}e^{-}$, pair annihilation, and Compton
scattering. GRBs resulting from NS--NS and NS--BH merging must be
accompanied by bursts of neutrino and gravitational radiation
(Eichler et al. 1989; Haensel et al. 1991; Narayan et al. 1992),
which could be possible targets for detection by existing
installations or those under construction. We note in passing
that large neutrino detectors and gravitational interferometers like
LIGO/VIRGO would give an opportunity to distinguish between the BH and NS
coalescence events because, in the case of BH merging, there will be a
relatively stronger gravitational and a less intense neutrino
radiation, compared to the merging of two NSs.

As it follows from the above considerations, there are potential tests
by which our cosmological scenario of GRB production could be 
observationally distinguished from the standard scenario. 

In this respect, detection of gravitational radiation from the vicinity
of a GRB would be decisive. The signature of the proposed scenario is the
specific pattern of gravitational wave radiation 
in the galactic nucleus associated with the origin of the given GRB.
The coalescence of the CSR binary, which is thought to result 
in the GRB, is accompanied by the presence
of {\it numerous} other close CSRs and their binaries
in the nucleus. Those objects are the sources
of excessive gravitational radiation, which continues to exist for
a long time after the gravitational radiation from the GRB disappears.
This differs drastically from the standard scenario, according to
which there are, along with the binary that has experienced 
coalescence, just a few (if any) binaries at a similar evolutionary stage
and thus no excessive gravitational radiation is expected after the GRB. 
In order to test our scenario, search for continuing gravitational radiation 
from the former GRBs could be one of the prime targets for LIGO/VIRGO
interferometer. 

Another possible signature of the evolved galactic nuclei scenario 
is a potential recurrency of GRBs. The more evolved the
galactic nucleus, the higher the possibility of finding there multiple
GRB events.  

The evolved galactic nuclei scenario for GRB origin outlined in this
paper makes it possible to draw the following conclusions:
\begin{description}
\item {(i)} The central parts
of the galactic nuclei, which in the course of the dynamical evolution
produce, through close encounters/collisions, coalescence, and SN
explosions of ordinary stars,
the abundant compact stellar remnants (NSs and BHs) as well as
binaries consisting of them,
could be the sites for the origin of GRB progenitors.
\item {(ii)} Actual location of a GRB and its afterglow might be far away
from the galactic nucleus (or even the parent galaxy) due to the ejection
of `superelastic' binaries as a result of interaction with the field
stars in the nucleus.
\item {(iii)} In contrast to the standard cosmological scenario of the GRB
origin, there is no need in {\it a priori} existence of a large number
of compact binaries consisting of NS. In compact galactic nuclei, random
encounters of CSR, which are accompanying by gravitational wave
radiation, result in the radiative binary formation and in further CSR
coalescence, and this would naturally explain
the GRB phenomenon.
\item {(iv)} The signature of the proposed scenario for the GRB origin
is the gravitational wave radiation that causes the coalescence of CSR
in the course of their close encounters. Spiralling in, which
accompanies the process of close binary formation and evolution, results
in a specific pattern of gravitational radiation, which is
distinguishable from the gravitational radiation of another origin.
\item {(v)} In principle, the recurrent GRBs (so far not yet detected with
certainty)
could be observable from a host galactic nucleus if it is far evolved.
\end{description}

If the proposed scenario for the origin of GRB progenitors in the evolved
galactic nuclei is confirmed by further observations, this would imply 
a major role played by CSR binary formation at the late stages of the 
dynamical evolution of galactic nuclei. If, on the other hand, it turns out
that the major fraction of GRB progenitors is associated, in its origin,
with the galactic disks, and not with the nuclei, it would lead to 
informative constraints on the evolutionary processes in the galactic nuclei.

\acknowledgments
We are grateful to V.S.~Beskin and G.V.~Chibisov for helpful
discussion and comments.
Yu.~E. acknowledges support of the Russian Fund for Fundamental Research 
under grants 96-15-96614 and 96-02-16670.

\end{document}